# Logical Inference by DNA Strand Algebra

KUMAR S. RAY
Electronics and Communication Science Unit
Indian Statistical Institute
203, B.T Road, Kolkata-700108, India
E-mail: ksray@isical.ac.in
Tel: +91 8981074174

MANDRITA MONDAL
Electronics and Communication Science Unit
Indian Statistical Institute
203, B.T Road, Kolkata-700108, India
E-mail: mandrita_1984@yahoo.co.in
Tel: +91 9830354798

**Abstract**

Based on the concept of DNA strand displacement and DNA strand algebra we have developed a method for logical inference which is not based on silicon based computing. Essentially, it is a paradigm shift from silicon to carbon. In this paper we have considered the inference mechanism, viz. modus ponens, to draw conclusion from any observed fact. Thus, the present approach to logical inference based on DNA strand algebra is basically an attempt to develop expert system design in the domain of DNA computing. We have illustrated our methodology with respect to worked out example. Our methodology is very flexible for implementation of different expert system applications.

**Keywords:** logical inference, DNA strand algebra; DNA gate; DNA signal; process algebra; DNA computing; modus ponens; strand displacement; molecular computing; expert system approach.

**1. Introduction**

Conventional silicon based computing has been successful over the past several decades. It is fast, flexible, and well understood. But it has reached its limitations of design complexity, processing power, memory, energy consumption, density and heat dissipation. Alternatives to these traditional methods of building computers have been proposed and in recent past molecular computing has gained considerable attention for non-silicon based computing which can overcome conventional computation problems. The Watson-Crick base pairing makes DNA a powerful tool for engineering at nano-scale [Adleman, 1994; Benenson et al., 2001; Green et al., 2006; Winfree et al., 1998]. The behaviour of DNA molecules can be predicted, and by setting up strands of DNA in the right way, they can draw conclusion from a set of logical inference rules [Ray and Mondal, 2011a; Ray and Mondal, 2011b].

In this paper we have performed logical reasoning by modus ponens using DNA strand algebra, which can be defined as a branch of process algebra. The operations of DNA strands in present model are based on the mechanism of DNA strand displacement. Process algebras are widely used for defining the formal semantics of concurrent communicating processes. Process calculi provide a tool for the high-level description of interactions, communications, and synchronizations between a collection of independent agents or processes. They also provide algebraic laws that allow process descriptions to be manipulated and analyzed, and permit formal reasoning about equivalences between processes. The main components of DNA strand algebra are DNA strands, DNA gates, and their interactions.

## 2. DNA strand algebra: syntax and semantics

Before going in detail description of DNA strand algebra we first have to know what is syntax and semantics. How are they different? [Cardelli, 2009; Cardelli 2013]

A language is a set of valid sentences. The validity of language can be broken down into two things: syntax and semantics. The term syntax refers to grammatical structure of a language, rather than what they refer to or mean. Whereas the term semantics is concerned to the meaning of the vocabulary symbols arranged with that structure, often in relation to their truth and falsehood. Grammatical (syntactically) valid does not imply sensible (semantically) valid. In mathematics, computer science and linguistics, a formal language is a set of strings of symbols that may be constrained by rules that are specific to it.

In computer science, the process algebras [Baeten, 2004] are mathematically rigorous languages with well defined semantics that permit describing and verifying properties of concurrent communicating systems. Process calculi provide a tool for the high-level description of interactions, communications, and synchronizations between a collection of independent agents or processes. They also provide algebraic laws that allow process descriptions to be manipulated and analyzed, and permit formal reasoning about equivalences between processes.

The chemistry of diluted well-mixed solutions where floating molecules can interact according to the reaction rules can be presented as process algebra [Berry et al., 1989]. Thus two binary relations can be defined on a set of chemical process algebra viz. *mixing* and *reaction*. Let, $P$, $Q$ and $R$ are three chemical solutions in a set A. Mixing ($P \equiv Q$) is an equivalence relation which syntactically brings the floating molecules close to each other so that they can interact. Reaction ($P \rightarrow Q$) describes how a solution becomes a different solution. The symmetric and transitive closure, $\rightarrow^*$ represents sequences of reactions. The chemical process algebra obeys the following general laws;

$P \equiv P$;  $P \equiv Q \Rightarrow Q \equiv P$;  $P \equiv Q, Q \equiv R \Rightarrow P \equiv R$ equivalence
$P \equiv Q \Rightarrow P + R \equiv Q + R$ in context

$P + Q \equiv Q + P; \quad P + (Q + R) \equiv (P + Q) + R; \quad P + 0 \equiv P$      diffusion
$P \rightarrow Q \Rightarrow P + R \rightarrow Q + R$      dilution
$P \equiv P', \ P' \rightarrow Q', \ Q' \equiv Q \Rightarrow P \rightarrow Q$      well mixing

*DNA strand algebra* can be defined as a branch of process algebra where the main components are DNA strands, DNA gates, and their interactions [Cardelli, 2009; Cardelli 2013]. Despite of similarity between DNA strand algebra and chemical reactions, or Petri nets, or multiset rewriting systems, there is a difference. The reaction, transition, and rewrite mechanisms of DNA algebra do not live *outside* the system, but rather are part of the system itself and are *consumed* by their own activity which reflects their DNA implementation.

The basic structure of DNA strand algebra contains atomic elements i.e. *signals* and *gates*, and two combinators: *parallel (concurrent) composition P | Q*, and *populations $P^*$*. An inexhaustible population $P^*$ has the property that $P^* = P | P^*$; that is, there is always one more $P$ that can be taken from the population. Signals are single stranded short DNA oligonucleotides. In this section, we represent signal strand as $x$.

A gate is an operator which absorbs a signal to produce one or more signals; sometimes produce nothing (0). Gates are generally single stranded or partially double stranded DNA molecules. Inert component 0, parallel compositions $P_1 | P_2$ and inexhaustible populations $P^*$ are assemblies of signals and gates. Gates and signals are combined in a 'soup' so that they floating molecules can interact. Signals can interact with gates but signals cannot interact with signals, nor gates with gates.

The input part of gate is Watson-Crick complementary with the corresponding signal strand. Let a DNA strand has three segments, viz. $b$, $c$, $d$. The *Watson-Crick complement* of this strand is $(b, c, d)^\perp = d^\perp, c^\perp, b^\perp$. The said DNA strand hybridized with its complementary strand i.e. the double stranded DNA sequence is shown in Fig. 1.

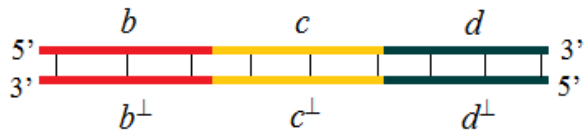

Figure 1. Double stranded DNA sequence

For example, $[x_1, \ldots, x_n].[x'_1, \ldots, x'_m]$ is a gate that binds signals $x_1, \ldots, x_n$ to produce signals $x'_1, \ldots, x'_m$, and is consumed in the process. Here the gate *joins* $n$ signals and then *forks* $m$ signals. Explanation of syntax and abbreviation is given by table 1.

| | |
|---|---|
| X | signal |
| 0 | inert |
| $P_1 | P_2$ | parallel compositions |
| $P^*$ | unbounded population |
| $x_1 \cdot x_2 \overset{def}{=} [x_1] \cdot [x_2]$ | transducer gate |
| $x \cdot [x_1, \ldots, x_m] \overset{def}{=} [x] \cdot [x_1, \ldots, x_n]$ | fork gate |
| $[x_1, \ldots, x_n] \cdot x \overset{def}{=} [x_1, \ldots, x_n] \cdot [x]$ | join gate |

Table 1. Syntax and abbreviation of DNA strand algebra

DNA strand algebra obeys the following laws for the binary relation *mixing* [Cardelli, 2009].

$$\left. \begin{array}{r} P \equiv P \\ P \equiv Q \Rightarrow Q \equiv P \\ P \equiv Q, Q \equiv P \Rightarrow P \equiv R \end{array} \right\} \text{equivalence}$$

$$\left. \begin{array}{r} P \mid 0 \equiv P \\ P \mid Q \Rightarrow Q \mid P \\ P \mid (Q \mid R) \equiv (P \mid Q) \mid R \end{array} \right\} \text{diffusion}$$

$$\left. \begin{array}{r} P \equiv Q \Rightarrow P \mid R \equiv Q \mid R \\ P \equiv Q \Rightarrow P^* \equiv Q^* \end{array} \right\} \text{in context}$$

$$\left. \begin{array}{r} P^* \equiv P^* \mid P \\ 0^* \equiv 0 \\ (P \mid Q)^* \equiv P^* \mid Q^* \\ P^{**} \equiv P^* \end{array} \right\} \text{population}$$

DNA strand algebra follows the following laws for the binary relation *reaction* [Cardelli, 2009].

| | |
|---|---|
| $x_1 \mid \ldots \mid x_n \mid [x_1, \ldots, x_n] \cdot [x'_1, \ldots, x'_m] \to x'_1 \mid \ldots \mid x'_m$ | gate ($n \geq 1, m \geq 0$) |
| $P \to Q \Rightarrow P \mid R \to Q \mid R$ | dilution |
| $P \equiv P', \ P' \to Q', \ Q' \equiv Q \Rightarrow P \to Q$ | well mixing |

Generally, DNA signal strand $x$ has three segments $x_h$, $x_t$, $x_b$ (in Fig. 3, $x$ is the signal strand containing three segments): $x_h$ = *history*, $x_t$ = *toehold*, $x_b$ = *binding*. The $x_h$ segment accumulate the history of previous interactions (it might even be hybridized). This segment is not part of signal identity. The $x_b$ segment hybridizes with the gate. A toehold segment, $x_t$, can reversibly interact with a gate and leads to toehold mediated branch migration and strand displacement [Cardelli, 2013; Zhang and Winfree 2009; Green and Tibbetts, 1981]. Strand displacement is the process through which two strands with partial or full complementarity hybridize to each other, displacing one or more pre-hybridized strands. Branch migration is the process by which a DNA sequence, partially paired to its complement in a DNA duplex, extends its pairing by displacing its resident strand. Strand displacement can be initiated at complementary single stranded segments, referred as toehold, and progress through a branch migration process. The pictorial representation of strand displacement is given below.

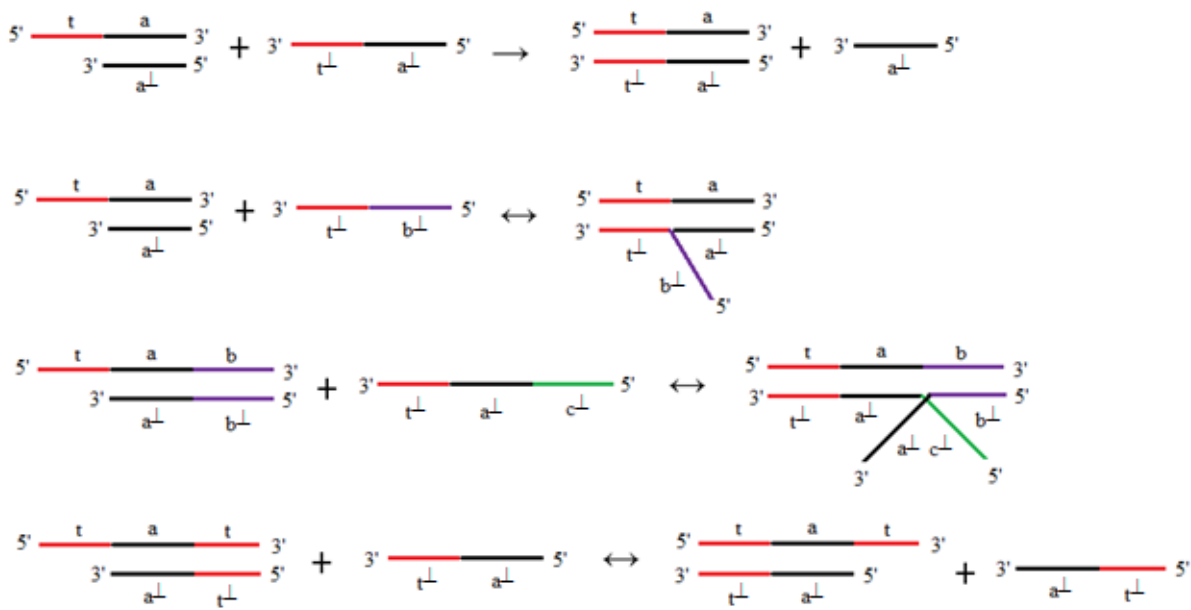

Figure 2. Toehold mediated DNA branch migration and strand displacement

In Fig. 2 each letter, i.e. $t$, $a$, $b$, $c$, represents a specific DNA oligonucleotide and their complementary sequences are shown by $t^\perp$, $a^\perp$, $b^\perp$, $c^\perp$ respectively. Each DNA strand is the concatenation of multiple oligonucleotides as shown in the above figure. The short sequences hybridize reversibly to their complementary sequences, whereas the hybridization of long sequences are irreversible; the exact critical length depends on the physical condition. Distinct letters indicate oligonucleotides that do not hybridize with each other.

In the first reaction of Fig. 2, a single stranded DNA sequence is mixed with a partial double stranded sequence. The toehold oligonucleotide $t$ initiates binding between the partial double stranded and the input single stranded sequence. After the hybridization of $t^\perp$, the $a^\perp$ part of the single stranded sequences gradually replaces the $a^\perp$ strand of the partially double stranded sequence by branch migration and finally displaces it. Thus after the reaction a fully double stranded sequence is formed releasing $a^\perp$ strand. This displacement reaction is irreversible because there is no toehold for the reverse reaction.

In the second reaction of Fig. 2 the input single stranded DNA sequence ($t^\perp b^\perp$) is partially complementary to the partial double stranded sequence. Thus, without replacing the other strand, the input sequence partially hybridizes reversibly with the double stranded sequence.

In the third reaction of Fig. 2, the input single stranded sequence can hybridize with the partially double stranded sequence up to a certain point and then reverts back to the toehold. Thus, no displacement occurs in this reaction.

In the fourth reaction of Fig. 2, shows toehold exchange. The input single stranded sequence, i.e. $t^{\perp}a^{\perp}$ displaces the sequence $a^{\perp}t^{\perp}$ of the partially double stranded sequence by branch migration. This reaction is reversible because of reverse toehold binding and branch migration.

In Fig. 3 represents the mechanism of annihilator [Cardelli, 2009]. Here, a gate $G$ binds to the signal strand $x$ by toehold mediated branch migration process. This reaction produces nothing (0). The gate can be called '$x_h$ generic' which means that the performance of the gate does not depend on $x_h$. This mechanism can be represented by the expression $x \mid x.[h] \rightarrow 0$.

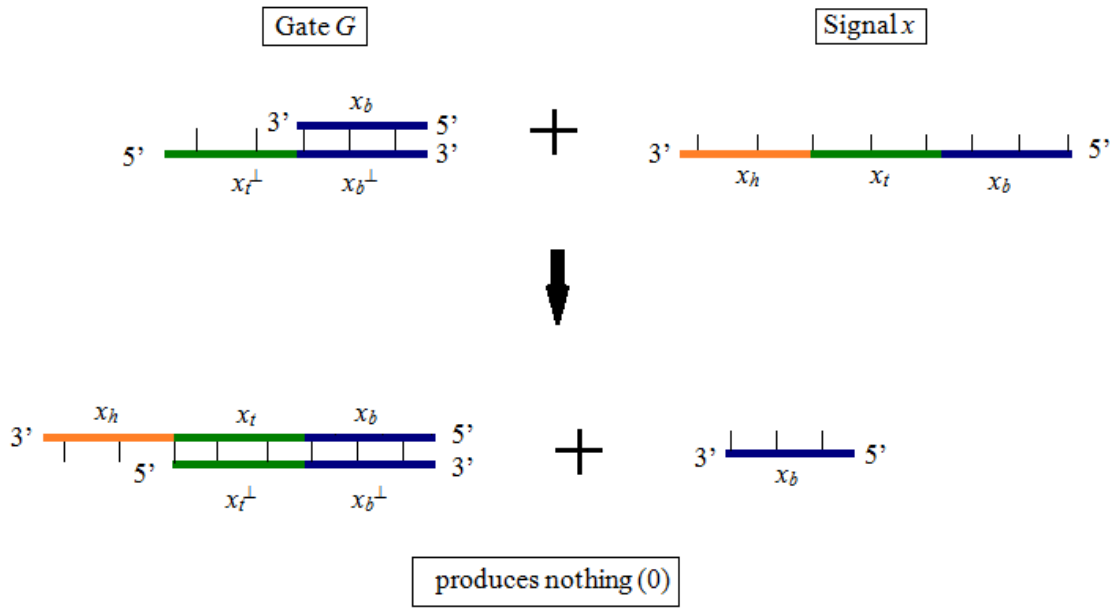

Figure 3. Annihilator

Fig. 4 represents the mechanism of transduer i.e. the gate $x.y$ which transduces a signal $x$ into a signal $y$. The gate works by two reactions, one is reversible and the other is irreversible. To perform these reactions there are two separate structures $G_b$ (gate backbone) and $G_t$ (gate trigger) of the gate. In the forward reaction, which is reversible, the signal $x$ hybridizes to $G_b$ and replaces the signal $y$ by strand displacement mechanism. The second $G_t$ reaction is irreversible as it 'locks' the gate in the state where the signal $x$ is consumed and $y$ is produced. This reaction produces nothing (0). This mechanism can be represented by the expression $x \mid x.y \rightarrow y$.

Generally, all the gate segments are fresh i.e. they are not shared by any other gate in the system.

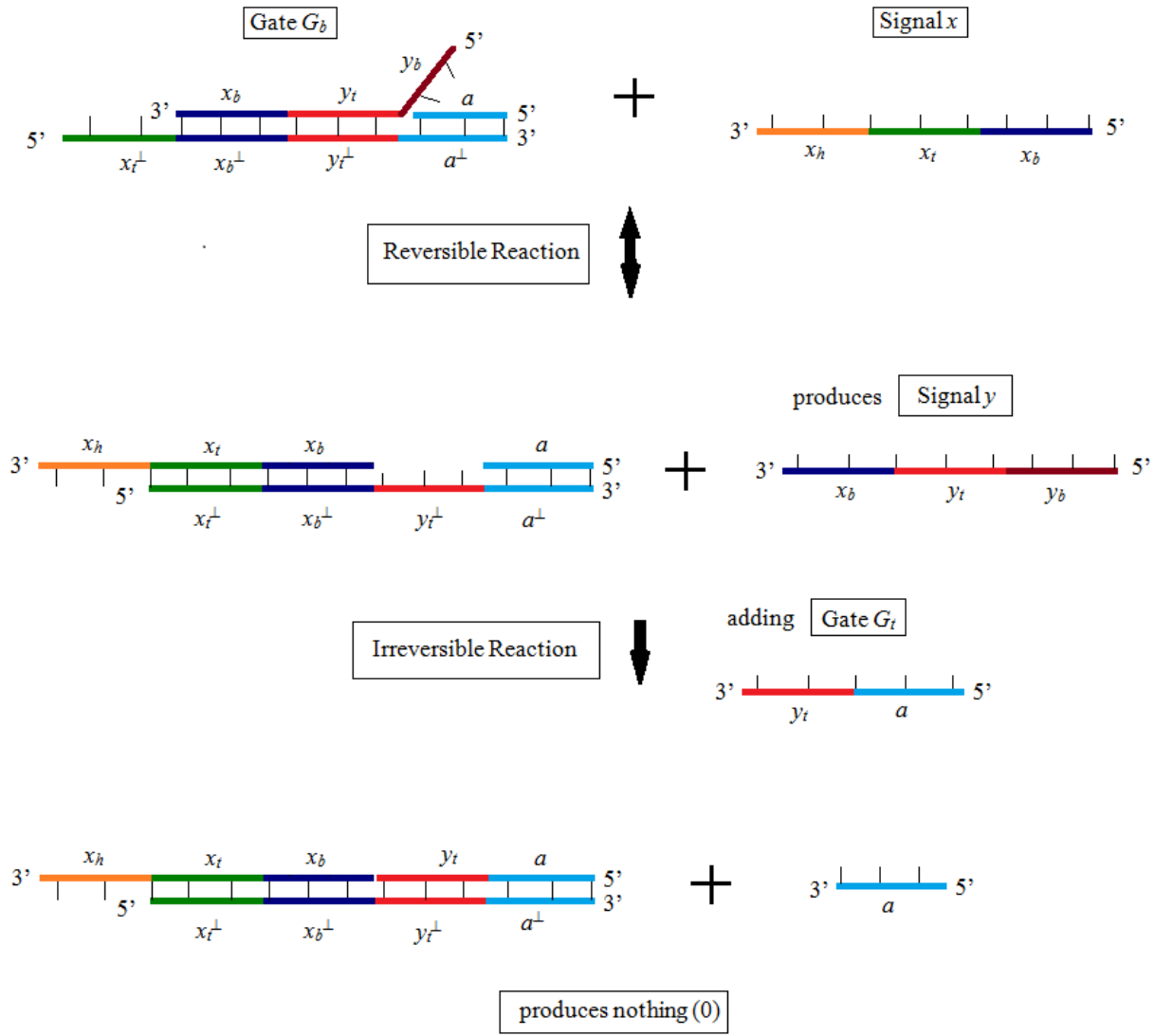

Figure 4. Transducer

Likewise, a transducer to a 2-way fork gate, $x.[y,z]$ which produces two output signals can also be formed. The expression of this system is $x \mid x.[y,z] \rightarrow y \mid z$. This can be extended to $n$-way fork, via longer trigger strands.

In his paper [Cardelli, 2009] Cardelli also shows the join gates by DNA strand algebra.

There is an extension of DNA strand algebra which is termed as *curried gates* i.e. gates that produce gates. For example, if $H(y)$ is a transducer $y.z$ as in Fig. 4, we obtain a curried gate $x.y.z$ such that $x \mid x.y.z \rightarrow y.z$.

## 3. Modus Ponens

In propositional logic modus ponens is an inference mechanism for two value based exact reasoning. Modus ponens law is formulated as an If….Then rule and is applied to classical two valued logic. This law is given by

*Premise 1* : If *X* is *A* Then *Y* is *B*
*Premise 2* : If *X* is *A*
___________________________________________
*Consequence*: *Y* is *B*.

A generalized form modus ponens, as shown below, is several conditional propositions combined with else [Ray and Mondal, 2011a; Ray and Mondal, 2011b].

That is,
*Premise 1* :   If *X* is $A_1$ and *Y* is $B_1$ then *Z* is $C_1$ else
*Premise 2* :   If *X* is $A_2$ and *Y* is $B_2$ then *Z* is $C_2$ else
          .                                                                                                      (1)
          .
*Premise n* :   If *X* is $A_n$ and *Y* is $B_n$ then *Z* is $C_n$ else
*Premise n+1* : If *X* is *A'* and Y is *B'*
___________________________________________
*Consequence*: *Z* is *C'*.

## 4. Formulation of Logical Inference by DNA Strand Algebra

Based on the syntax and semantics of DNA strand algebra as discussed in section 2, we have designed DNA strand algebra that can perform logical inference using modus ponens.

We will deduce logical inference with the help of an example where we have two different domains in the antecedent parts and one domain in the consequent part. The two domains of antecedent part are height and weight, thus, we get two antecedent clauses in each rule which is stated by an expert. The one domain in the consequent part is body mass index. Hence we have one clause in the consequent part of each rule of our knowledgebase. Thus, the generalized form of modus ponens as stated in section 3 is now reduced to a specific form having a finite number of rules and each rule is having two antecedent clauses and one consequent clause. In this section we first consider the dynamic range of three domains (viz. height, weight and body mass index).

Let, the Universe of Height (domain *A*) is denoted by Ht, Universe of Weight (domain *B*) is denoted by Wt and the Universe of Body Mass Index (domain *C*) is denoted by BMI.

Quantization of Universes of Ht, Wt and BMI are considered in tables 1, 2 & 3. The corresponding quantized ranges are also represented linguistically and by DNA oligonucleotide sequences in the above said tables. The short DNA strands representing the quantized ranges are given in 5' to 3' direction. The complementary sequences in 3' to 5' direction of the short DNA strands also represent the corresponding quantized range. For example, the DNA strand 5'TAATT3' represents the quantized range 4'3" ≤ Ht ≤ 4'6" (linguistic value is very short (II)) of Ht domain. The complementary sequence 3'ATTAA5' also represents the same quantized range 4'3" ≤ Ht ≤ 4'6".

Table: 1  Quantization of Height (*A* Domain)

| Quantized Universe | Oligonucleotide Sequences (5'-3') | Linguistic Value |
|---|---|---|
| Ht < 4'3" | CTGGA | Very Short(I) |
| 4'3 ≤ Ht < 4'6" | TAATT | Very Short(II) |
| 4'6" ≤ Ht < 4'9" | GATCC | Short(I) |
| 4'9" ≤ Ht < 5' | ATTTT | Short(II) |
| 5' ≤ Ht < 5'3" | TCAGC | Medium Height(I) |
| 5'3" ≤ Ht < 5'6" | CGAAT | Medium Height(II) |
| 5'6" ≤ Ht < 5'9" | AATGT | Tall(I) |
| 5'9" ≤ Ht < 6' | CCGGA | Tall(II) |
| 6' ≤ Ht < 6'3" | ATCGT | Very Tall(I) |
| 6'3" ≤ Ht | TTAGA | Very Tall(II) |

Table: 2  Quantization of Weight (*B* Domain)

| Quantized Universe | Oligonucleotide Sequences (5'-3') | Linguistic Value |
|---|---|---|
| Wt < 90 lb | ATTCA | Very Light(I) |
| 90 lb ≤ Wt < 100 lb | GCCAA | Very Light(II) |
| 100 lb ≤ Wt < 110 lb | TTCGT | Light(I) |
| 110 lb ≤ Wt < 120 lb | CAAAC | Light(II) |
| 120 lb ≤ Wt < 130 lb | CGGAA | Medium Weight(I) |
| 130 lb ≤ Wt < 140 lb | ATCCG | Medium Weight(II) |

| 140 lb ≤ Wt < 150 lb | GGAAT | Heavy(I) |
| 150 lb ≤ Wt < 160 lb | GTAGC | Heavy(II) |
| 160 lb ≤ Wt < 170 lb | ATCCC | Very Heavy(I) |
| 170 lb ≤ Wt | TAGGA | Very Heavy(II) |

Table: 3  Quantization of Body Mass Index (*C* Domain)

| Quantized Universe | Oligonucleotide Sequences (5'-3') | Linguistic Value |
| --- | --- | --- |
| BMI < 18.5 | CTAAG | Under Weight |
| 18.5 ≤ BMI < 25 | AGGAA | Normal Weight |
| 25 ≤ BMI < 30 | TAGCT | Over Weight |
| 30 ≤ BMI < 35 | GCGCG | Obesity (Class I) |
| 35 ≤ BMI < 40 | GTAAC | Obesity (Class II) |
| 40 ≤ BMI | AAATA | Morbid Obesity |

The relation among domains *A*, *B* and *C* (i.e. Ht, Wt and BMI) is:

$$\text{BMI} = \left[\frac{\text{Wt in lb}}{(\text{Ht in inch})^2}\right] \times 703$$

The total number of rules in our knowledgebase is as follows:
>There are 10 linguistic values in *A* domain and 10 linguistic values in *B* domain.
>Therefore, total number of possible rules = (10 × 10) = 100

Few rules are stated below:

If $A_1 \rightarrow$ Medium Ht(I)  and $B_1 \rightarrow$ Very Light(II)  then $C_1 \rightarrow$ Under Wt
If $A_2 \rightarrow$ Short(I)  and $B_2 \rightarrow$ Very Heavy(I)  then $C_2 \rightarrow$ Obesity (Class II)
If $A_3 \rightarrow$ Very Tall(I)  and $B_3 \rightarrow$ Heavy(I)  then $C_3 \rightarrow$ Normal Wt
If $A_4 \rightarrow$ Very Short(I)  and $B_4 \rightarrow$ Heavy(II)  then $C_4 \rightarrow$ Morbid Obesity
If $A_5 \rightarrow$ Short(II)  and $B_5 \rightarrow$ Very Heavy(II)  then $C_5 \rightarrow$ Obesity (Class II)
If $A_6 \rightarrow$ Very Short(II)  and $B_6 \rightarrow$ Medium Wt(II)  then $C_6 \rightarrow$ Obesity (Class I)
If $A_7 \rightarrow$ Tall(I)  and $B_7 \rightarrow$ Very Heavy(I)  then $C_7 \rightarrow$ Over Wt
If $A_8 \rightarrow$ Medium Ht(I)  and $B_8 \rightarrow$ Heavy(I)  then $C_8 \rightarrow$ Over Wt
If $A_9 \rightarrow$ Very Tall(II)  and $B_9 \rightarrow$ Medium Wt(I)  then $C_9 \rightarrow$ Under Wt
If $A_{10} \rightarrow$ Very Short(II)  and $B_{10} \rightarrow$ Very Light(II)  then $C_{10} \rightarrow$ Normal Wt

## 4.1. Statement of the problem

Let, the knowledgebase with all the rules are given (see model of equation (1)). Now, the given problem is;

If Ht ($A'$) → Medium Ht(I) and Wt ($B'$) → Heavy(I) then $C'$ → ?

We have to search the knowledgebase to select the desired rule by exact matching of the antecedent clauses. The conclusion can be derived from the selected rule.

## 4.2. Designing gate backbone ($G_b$)

For the formulation of the problem by DNA strand algebra we have to design certain gate structure. We need two separate gate structures, gate backbone ($G_b$) and gate trigger ($G_t$). First, we will discuss the structure of $G_b$.

All the rules in the knowledgebase are encoded by DNA strands. Using these DNA strands the gate backbones ($G_b$) are formed. As we have 100 rules in our knowledgebase, so there will be 100 separate structures of gate backbones in the test tube containing the solution.

For the formation of $G_b$ we need two five base long DNA strands, one will indicate the starting point of two antecedent clauses and the other will indicate the starting point of the consequent clause. The first DNA strand AAAAA is termed as *ant* and the second DNA strand CCCCC is termed as *con*. These two domains are of gate structure are fixed for all the rules in the knowledgebase.

Let, we want to encode rule $x$ in form of $G_b$, where $x$ can be any rule of the knowledgebase. The pictorial representation the $G_b$ is Fig. 5.

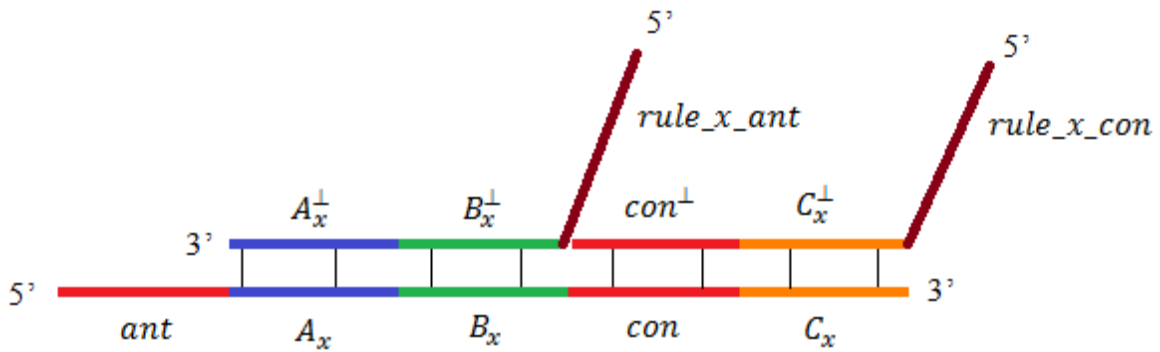

Figure 5. Gate backbone ($G_b$) representing rule $x$

In Fig. 5, $A_x$ represents the five base long DNA strand which encodes the Ht domain of rule $x$. Similarly, $B_x$ and $C_x$ encode the Wt and BMI domain respectively. $A_x$ and $B_x$ are the antecedent clause and $C_x$ is the consequent clause of rule $x$. These domains are the parts of the DNA strand which is in 5' to 3' direction. There are two different DNA strands in $G_b$ which are in 3' to 5' direction. The first strand, termed as *signal B*, contains the segments $A_x^\perp$ and $B_x^\perp$ which are complementary to $A_x$ and $B_x$ respectively. These complementary sequences represent the antecedent clause of the rule. The 5' end of *signal B* has a segment *rule_x_ant* which is 10 base long arbitrary DNA strand. This segment holds the rule number i.e. it indicates antecedent clause of the particular rule *x*, as $A_x^\perp$ and $B_x^\perp$ are attached to this strand. The segment *rule_x_ant* is different for each rule in the knowledgebase. Similarly, the second DNA strand of the gate backbone which is in 3' to 5' direction, termed as *signal C*, contains segments $con^\perp$ and $C_x^\perp$ which are complementary to *con* and $C_x$ respectively. The 5' end of *signal C* has a segment *rule_x_con* which is 10 base long arbitrary DNA strand. This segment indicates consequent clause of the particular rule *x* as *con* and $C_x$ are attached to this strand. The segment *rule_x_con* is different for each rule in the knowledgebase.

For example, we will show the formation of $G_b$ which encodes rule 1 i.e. "if $A_1 \rightarrow$ Medium Ht(I) and $B_1 \rightarrow$ Very Light(II) then $C_1 \rightarrow$ Under Wt".

Thus, for $x = 1$, $A_x = A_1 =$ Medium Ht(I), $B_x = B_1 =$ Very Light(II) and $C_x = C_1 =$ Under Wt.

The DNA strand which is in 5' to 3' direction in the gate backbone is shown below (refer table 1, 2 and 3):

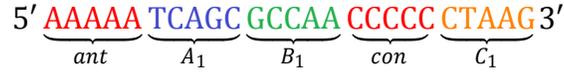

*Signal B* which is in 3' to 5' direction in the gate backbone is shown below. The DNA strand encoding *rule_1_ant* segment is unique for rule 1.

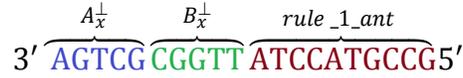

*Signal C* which is in 3' to 5' direction in the gate structure is shown below. The DNA strand encoding *rule_1_con* segment is unique for rule 1.

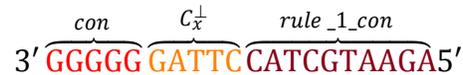

The gate backbone ($G_b$) encoding rule 1 is shown in Fig. 6. The strand in 5' to 3' direction is continuous. There is no nick in this strand. But the complementary strand in 3' to 5' direction consists of two different DNA sequences as stated earlier. One is *signal B* and the other is *signal C*.

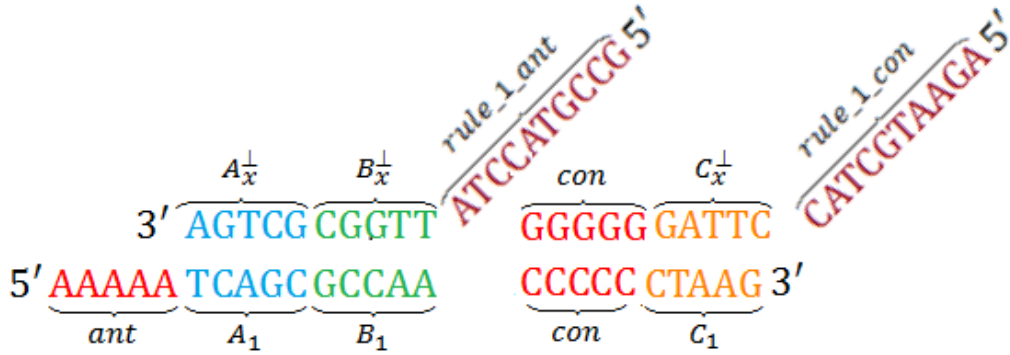

Figure 6. $G_b$ representing rule 1

### 4.3. Designing signal strand

To solve the problem of logical inference by DNA strand algebra, signal strand is needed to hybridize with the gate. The given observed data (section 4.1) is coded by single stranded DNA sequence which is presented as signal. The first signal strand is termed as *signal A* which contains the observed Ht domain of the antecedent clause i.e. *A'*. Signal A has three segments (see Fig. 7). The segment at 3' end is *obs*. It is 10 base long arbitrarily chosen DNA strand which is unique for any observed data. The segment next to *obs* is $ant^\perp$. It is complementary to the DNA strand *ant*. Thus, the DNA sequence representing $ant^\perp$ is TTTTT. The next segment i.e. the segment at 5' end represents the observed data *A'*. According to the given problem (section 4.1), *A'* is medium ht(I). The sequence coding medium ht(I) is TCAGC (table 1). As *signal A* is in 3' to 5' direction, the DNA strand encoding medium ht(I) in the signal is complementary to TCAGC i.e. AGTCG. *Signal A* is shown in Fig. 7.

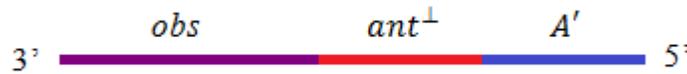

Figure 7. *Signal A*

### 4.4. Designing gate trigger ($G_t$)

For formulation of the problem, another gate structure is needed which is termed as gate trigger ($G_t$). $G_t$ has three segments (see Fig. 8). The first segment at 3' end represents the observed data *B'*. According to the given problem (section 4.1), *B'* is heavy(I). The sequence coding heavy(I) is GGAAT (table 2). As $G_t$ is in 3' to 5' direction, the DNA strand encoding heavy(I) is complementary to GGAAT i.e. CCTTA. The segment next to *B'* is complementary to the DNA strand *con*. Thus, the DNA sequence representing $con^\perp$ is GGGGG. The segment at 5' end of $G_t$ represents the possible consequence of the given antecedent clause. This segment is termed as $C'_y$, where *y* is under weight or normal weight or over weight or obesity (class I) or obesity (class II) or morbid obesity. $C'_y$ is complementary strand to the corresponding consequence.

For example, let us consider *y* is under weight. The short DNA strand representing under weight is CTAAG. Thus, $C'_{under\ wt}$ is the DNA sequence GATTC. Similarly, if *y* is obesity (class II), the short DNA strand representing $C'_{obesity\ (II)}$ is the DNA sequence CATTG (as obesity (class II) is encoded by GTAAC).

So, there should be six different structures of $G_t$ containing all possible consequences of the observed data. The structure of gate trigger ($G_t$) is shown in Fig. 8.

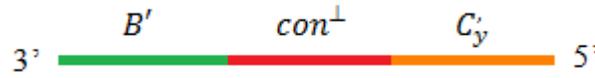

Figure 8. Structure of gate trigger ($G_t$)

## 4.5. Algorithm

Step 1:

All the rules in the knowledgebase are coded in the form of gate backbone ($G_b$) as shown in section 4.2. The gates are partially double stranded structure. Thus, there should be 100 of separate structures of $G_b$ in the test tube as the knowledgebase contains 100 rules.

Step 2:

The designed *signal A* is added in the test tube containing the solution which combines to the desired gate backbone ($G_b$). The segment $ant^\perp$ of *signal A* is the toehold which hybridizes to *ant* segment of desired $G_b$ and gradually the hybridization progress through branch migration. The segment *ant* is single stranded overhang at 5' end of the gate backbone. Finally displacing *signal B* (containing the segments $A_x^\perp$, $B_x^\perp$ and *rule_x_ant*), *signal A* hybridizes to $G_b$. This reaction is reversible. The mechanism is shown in Fig. 9.

Among 100 separate structures of $G_b$, *signal A* chooses the particular gate backbone by exact matching whose $A_x$ segment is Watson-Crick complement of *A'* (segment of *signal A*).

Step 3:

All DNA strands representing $G_t$ are added to the solution. There should be six different structures of $G_t$. Among all these only a single strand can bind to the gate derived from step 2. *B'* segment of the particular strand representing $G_t$ is the toehold which is Watson-Crick complement to $B_x$ segment. If the remaining segments of $G_t$ are also complementary to the gate derived from step 2, then only $G_t$ can hybridize via toehold mediated branch migration mechanism. $G_t$ replaces *signal C* and completely binds to the gate as shown in Fig. 9. This reaction is irreversible and it locks the gate as there is no open toehold. The gate is now in inert state.

The mechanism shown in Fig. 9 can be represented by the following expression:
$$A \,|\, A\,.\,[B,C] \rightarrow B\,|\,C$$
This gate structure consumes one input *signal A* and produces two outputs *signal B* and *signal C*. *Signal C* is the carries the conclusion of the given observation.

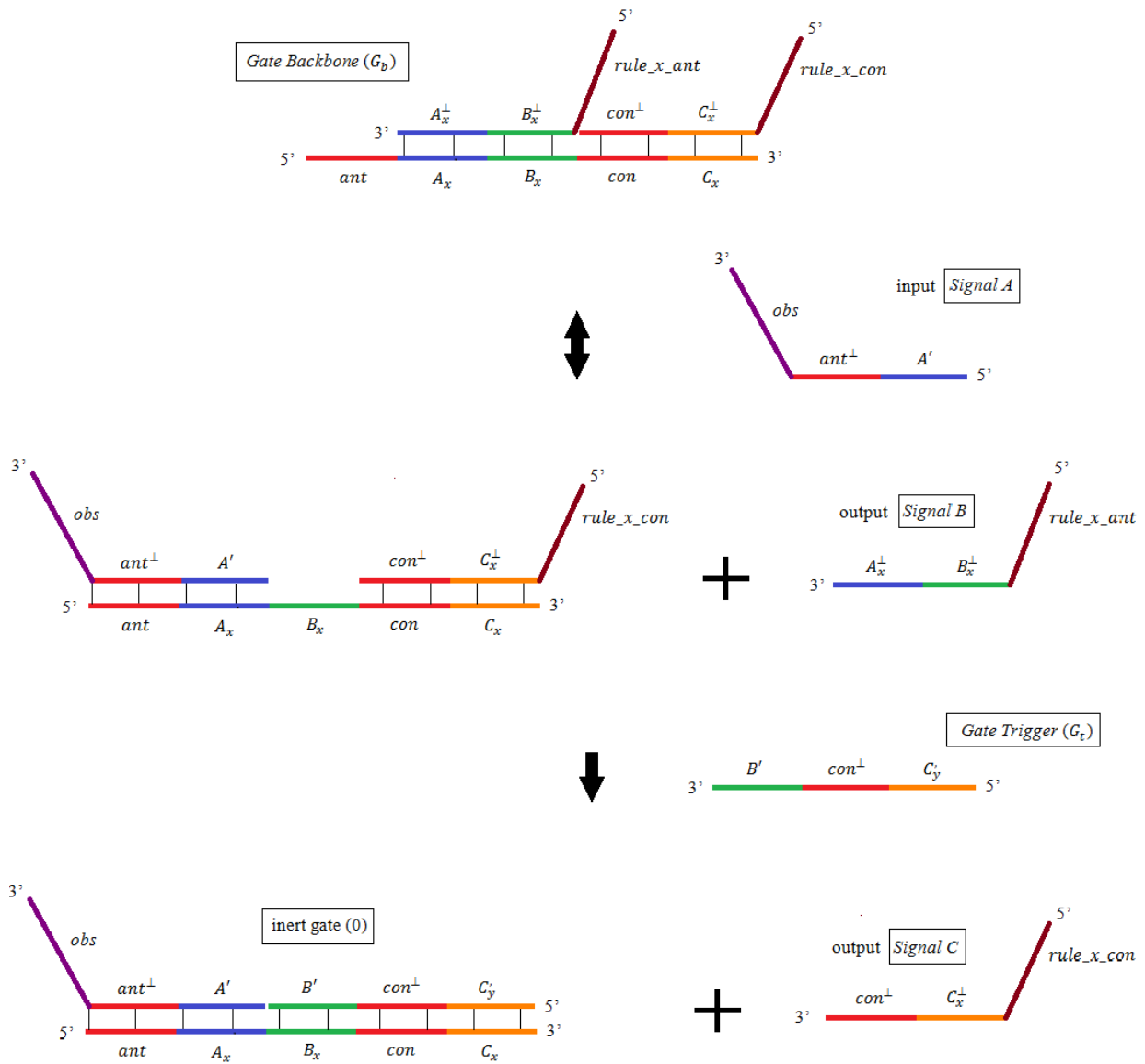

Figure 9. Pictorial representation of deduction of logical inference

Step 4:

In the test tube, there are 100 separate structures of $G_b$. Thus, 100 separate structures of *signal C* are present. To recover our desired *signal C* which is displaced by toehold mediated branch migration, all the Watson-Crick complements of *signal C* are added. One of these complementary sequences hybridizes with displaced single stranded *signal C* to form a complete double stranded end product. The hybridization is shown in Fig. 10.

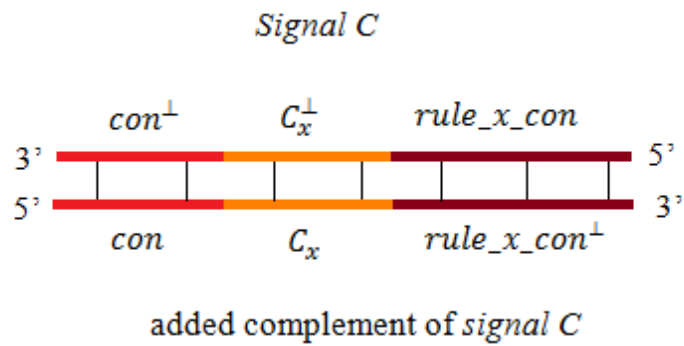

Figure 10. Production of complete double stranded end product

Step 5:

The solution of the test tube is divided in two sample test tubes. The first sample is treated with exonuclease enzyme and the second sample is not treated with any enzyme. The treatment with exonuclease causes degradation of single stranded DNA sequences including sticky ends, or double strands which are not completely hybridized. The degraded sequences are smaller fragments or single nucleotides.

Step 6:

Gel electrophoresis is performed with two DNA samples. By comparing the electrophoretograms for both DNA samples we draw final conclusion on existence or not existence of the desired double stranded molecule in the reaction test tube. On both electrophoretograms the location of at least one DNA band is not changed. This band is for the completely hybridized double stranded DNA sequence which has blunt ends. This complete double stranded sequence is the end product i.e. double stranded *signal C*.

Step 7:

The order of the bases of the desired double stranded sequence is known from sequencer. The conclusion of the given observation can be known from the resultant sequence.

## 5. Result and discussion

The complete double stranded end product i.e. *signal C* contains three segments. The first segment $con^\perp$ GGGGG (or *con*) represents the starting point of the consequent clause. The next segment $C_x^\perp$ represents the inferred conclusion of the given observation. By analyzing the reading of sequencer it can be shown that $C_x^\perp$ is ATCGA. It is Watson-Crick complement of TAGCT which represents the linguistic value 'over weight' of the domain BMI. The segment at 5' end of *signal C* is *rule_x_con* which holds the rule number. This segment indicates the particular rule of the knowledgebase which is selected to infer the conclusion of given observed data.

The observed data which we have considered in this paper is Ht (*A'*) → medium ht(I) and Wt (*B'*) → heavy(I). For this case rule 8 is selected from the knowledgebase. The conclusion (*C'*) 'over weight' is inferred from this rule. Thus, we can say if height is medium ht(I) and weight is heavy(I) then BMI is over weight. The segment at 5' end of *signal C* is the 10 bases long DNA strand *rule_8_con*. This strand is unique for rule 8 and represents that rule 8 is selected from the knowledgebase.

## 6. Conclusion

We have successfully developed one method for logical inference based on DNA strand algebra. Note that, by the word 'logical inference' we essentially mean to implement modus ponens which is a familiar and well established inference mechanism. Instead of using any notion of classical logic in terms of syntax and semantics we consider the chemical potentials of DNA strands. We have exploited the power of DNA strand displacement and flexibility of DNA strand algebra which is essentially derived from process algebra. Thus, an inference mechanism has been implemented based on DNA chemistry. We have explained and tested our methodology on detail worked out example. The merit of this mechanism can be further extended for implementation of rule based expert system design. Thus a new approach to DNA computing has been established.

**References:**

Adleman, L. (1994) 'Molecular computation of solutions to combinatorial problems', Science, Vol. 266, No. 5187, pp.1021–1024.

Baeten J.C.M., (2004). "A brief history of process algebra". Rapport CSR 04-02 (Vakgroep Informatica, Technische Universiteit Eindhoven).

Benenson Y., Paz-Elizur T., Adar R., Keinan E., Livneh Z., and Shapiro E., (2001) 'Programmable and autonomous computing machine made of biomolecules', Nature, Vol. 414, No. 22, pp.430–434.

Berry G., Boudol G., (1989), "The Chemical Abstract Machine", Proc. 17th POPL, ACM, 81-94.

Cardelli L., (2009), "Strand Algebras for DNA Computing", DNA Computing and Molecular Programming: 15th International Conference, DNA 15, Fayetteville, AR, USA, June 8-11, 2009, Revised Selected Papers, Springer-Verlag, Berlin, Heidelberg, 2009.

Cardelli L., (2010) Proc. 6th Workshop Developments in Computational Models, edited by S.B. Cooper, E. Kashefi and P. Panangaden, Electr. Proc. Theor. Comput. Sci., Vol. 26, pp.33–47.

Cardelli L., (2013) "Two-domain DNA strand displacement", Mathematical Structures in Computer Science 23(2): 247-27.

Green C. and Tibbetts C. (1981) 'Reassociation rate limited displacement of DNA strands by branch migration', Nucleic Acids Res., Vol. 9, No. 8, pp.1905–1918.

Green S.J., Lubrich D., Turberfield A.J., (2006) 'DNA hairpins: fuel for autonomous DNA devices', Biophysical Journal, Vol. 91, No. 8, pp.2966–2975.

Ray K.S. and Mondal M., (2011a), 'Similarity-based fuzzy reasoning by DNA computing', International Journal of Bio-inspired Computation, Vol. 3, No. 2, pp.112–122.

Ray K.S. and Mondal M., (2011b), 'Classification of SODAR data by DNA computing', New Mathematics and Natural Computation, Vol. 7, No. 3, pp.413–432.

Winfree E., Liu F., Wenzler L.A., and Seeman N.C., (1998) 'Design and self assembly of two dimensional DNA crystals', Nature, Vol. 394, No. 6693, pp.539–544.

Zhang D.Y. and Winfree E., (2009) 'Control of DNA strand displacement kinetics using toehold exchange', Journal of American Chemical Society, Vol. 131, No. 47, pp.17303–17314.